\documentstyle[epsfig,psfig,12pt,aaspp4]{article}
\lefthead{Naik, Paul and Callanan}
\righthead{GX~1+4 with {\it BeppoSAX}}
\slugcomment{Accepted for publication in The Astrophysical Journal}
\newcommand\sax{{\it BeppoSAX}}
\newcommand\gx{GX~1+4}

\received{}
\begin{document}
\title{Detection of hard X-ray pulsations and a strong iron $K_\beta$ emission 
line during an extended low state of GX~1+4}

\author{S. Naik\altaffilmark{1}, B. Paul\altaffilmark{2} and P. J. Callanan\altaffilmark{1}}
\altaffiltext{1}{Department of Physics, University College Cork, Cork, Ireland,~sachi@ucc.ie, paulc@miranda.ucc.ie}
\altaffiltext{2}{Tata Institute of Fundamental Research, Homi Bhabha Road, Mumbai 400 005, India,~bpaul@tifr.res.in}

\begin {abstract}
We present here results obtained from a detailed timing and spectral 
analysis of three \sax\ observations of the binary X-ray pulsar \gx\ 
carried out in August 1996, March 1997, and August 2000. In the middle 
of the August 2000 observation, the source was in a rare low intensity 
state that lasted for about 30 hours. Though the source does not show 
pulsations in the soft X-ray band (1.0--5.5 keV) during the extended 
low state, pulsations are detected in 5.5--10.0 keV energy band of the 
MECS detector and in hard X-ray energy bands (15--150 keV) of the PDS
instrument. Comparing the 2--10 keV flux during this low state with 
the previously reported low states in GX 1+4, we suggest that the 
propeller regime in GX~1+4 occurs at a lower mass accretion rate than 
reported earlier. Broad-band (1.0--150 keV) pulse averaged spectroscopy 
reveals that the best-fit model comprises of a Comptonized continuum 
along with an iron $K_\alpha$ emission line. A strong iron $K_\beta$ 
emission line is detected for the first time in \gx\ during the extended 
low state of 2000 observation with equivalent width of $\sim$ 550 eV. 
The optical depth and temperature of the Comptonizing plasma are found to
be identical during the high and low intensity states whereas the hydrogen 
column density and the temperature of the seed photons are higher during 
the low state. We also present results from pulse phase resolved
spectroscopy during the high and low flux episodes.
\end{abstract}


\section{Introduction}

The luminous accretion-powered X-ray pulsar GX~1+4 has several unique 
characteristics which make it an ideal source for a detailed study in a
wide X-ray energy band. The optical counterpart is a M5 III giant star
in a rare type of symbiotic system (Chakrabarty \& Roche 1997). The neutron
star in the binary system is a slow pulsar with a spin period of about 2 
minutes. It is one of the brightest and hardest X-ray sources in the sky 
with a large rate of change of pulse period. In 1970s, the pulsar 
exhibited spin-up behavior and after an extended low intensity state in 
early 1980s, it showed spin-down activity (Makishima et al. 1988).
$BATSE$ monitoring of the source, since 1991, confirmed the spin down trend
with occasional torque reversal events (Chakrabarty et al. 1997). 
Though the spin-up torque is expected to be correlated with the
X-ray luminosity (Ghosh \& Lamb 1979), they have been found instead 
to be anticorrelated for \gx\ (Chakrabarty et al. 1997, Paul et al. 1997a).
The pulsed X-ray luminosity of \gx\ monitored with $BATSE$, and the total 
X-ray luminosity monitored with RXTE-ASM, shows strong variability over days 
to years time scale, but without any periodic nature.

The pulse profile of \gx\ has a characteristic dip, which has been
attributed due to accretion column eclipses (Galloway et al. 2001).
The dip is broad when the source is bright and it becomes narrow with
decreasing intensity. In the hard X-rays, a change in pulse profile, from a
single peak to a double peak with associated flux change was noted in \gx\
with balloon borne observations (Rao et al. 1994, Paul et al. 1997b).
It occasionally shows a low state of a few hours duration ($\sim$ 40 ks
in July 1996) during which the pulsations are absent (Cui 1997, Cui and
Smith 2004). This has been interpreted as due to centrifugal prohibition
of accretion, also known as the `propeller effect'. As \gx\ is spinning down
in spite of a large accretion rate (the persistent X-ray luminosity is a 
few times 10$^{37}$ erg s$^{-1}$), the magnetic field strength of the 
neutron star is expected to be high, 10$^{13-14}$ G (Makishima et al.
1989) although this is yet to be confirmed from spectroscopic measurements. 
Based on the fluctuation of the spin-up rate, Pereira et al. (1999) suggested
a binary period of 304 days.

The continuum energy spectrum of accreting pulsars is generally described 
by a model consisting of a power-law, which is known to be a signature of 
unsaturated Comptonization and a Gaussian function for the Iron K$_\alpha$ 
emission line feature. Iron K shell emission lines in X-ray pulsars are 
believed to be produced by illumination of neutral or partially ionized 
material in the accretion disk, stellar wind of the high mass companion, 
material in the line of sight, or in the accretion column. The spectral 
fitting to the phase averaged $RXTE$ data, however, shows that the GX~1+4 
energy spectrum is best fitted by a model consisting of an analytic 
approximation to a Comptonization continuum component and a Gaussian 
component, attenuated by the neutral absorption column density (Galloway 
2000). Pulse-phase resolved spectroscopy shows a significant increase in 
the value of optical depth $\tau$ during the dips. This reveals that the 
dip features in the pulse profiles are due to the eclipses of the emitting 
region by the accretion column (Galloway et al. 2000).

In this paper, we present the results obtained from a detailed timing and 
spectral analysis of three observations of \gx, made in August 1996,
March 1997, and August 2000, with the Low Energy Concentrator Spectrometer
(LECS), Medium Energy Concentrator Spectrometers (MECS), and the hard X-ray
Phoswich Detector System (PDS) instruments of \sax\ in 1.0--150.0 keV
energy band. During a major part of the August 2000 observation, the X-ray
flux was at a very low level, similar to a state previously seen with
$RXTE$ in 1996 July, but for a longer duration of about 30 hr. The \sax\
observation provides us with an opportunity to investigate the temporal and 
spectral properties of \gx\ during this extended low state over a much wider
energy band and better energy resolution than before. In subsequent 
sections we give details of the three \sax\ observations, the results obtained 
from the timing and spectral analysis, followed by a discussion of the results.

\section{Observations}

The observations of \gx, used for the present study, were carried out with
LECS, MECS, and PDS instruments of the \sax\ satellite in 1996 August, 1997 
March, and 2000 August. The details of the observations with useful exposure 
times are given in Table~\ref{obs}. The MECS consists of three grazing
incidence telescopes with imaging gas scintillation proportional counters in
their focal planes. The LECS uses an identical concentrator system as the MECS, 
but utilizes an ultra-thin entrance window and a driftless configuration to 
extend the low-energy response to 0.1 keV. The PDS detector system is composed 
of 4 actively shielded NaI(Tl)/CsI(Na) phoswich scintillators with a total 
geometric area of 795 cm$^2$ and a field of view of 1.3$^o$ (FWHM).
The LECS, MECS and PDS are sensitive in the energy bands
of 0.1-5.0, 1.0-10.0 and 15--300 keV respectively. The energy resolutions
of LECS, MECS, and PDS are 25\% at 0.6 keV, 8\% at 6 keV and $\leq$ 15\% at
60 keV respectively. Time resolution of the instruments during these 
observations was 15.25 $\mu$s. For a detailed description of the \sax\ 
mission, refer to Boella et al. (1997) and Frontera et al. (1997).

\section{Timing Analysis}

\begin{figure}
\centering
\vskip 9.2cm
\includegraphics{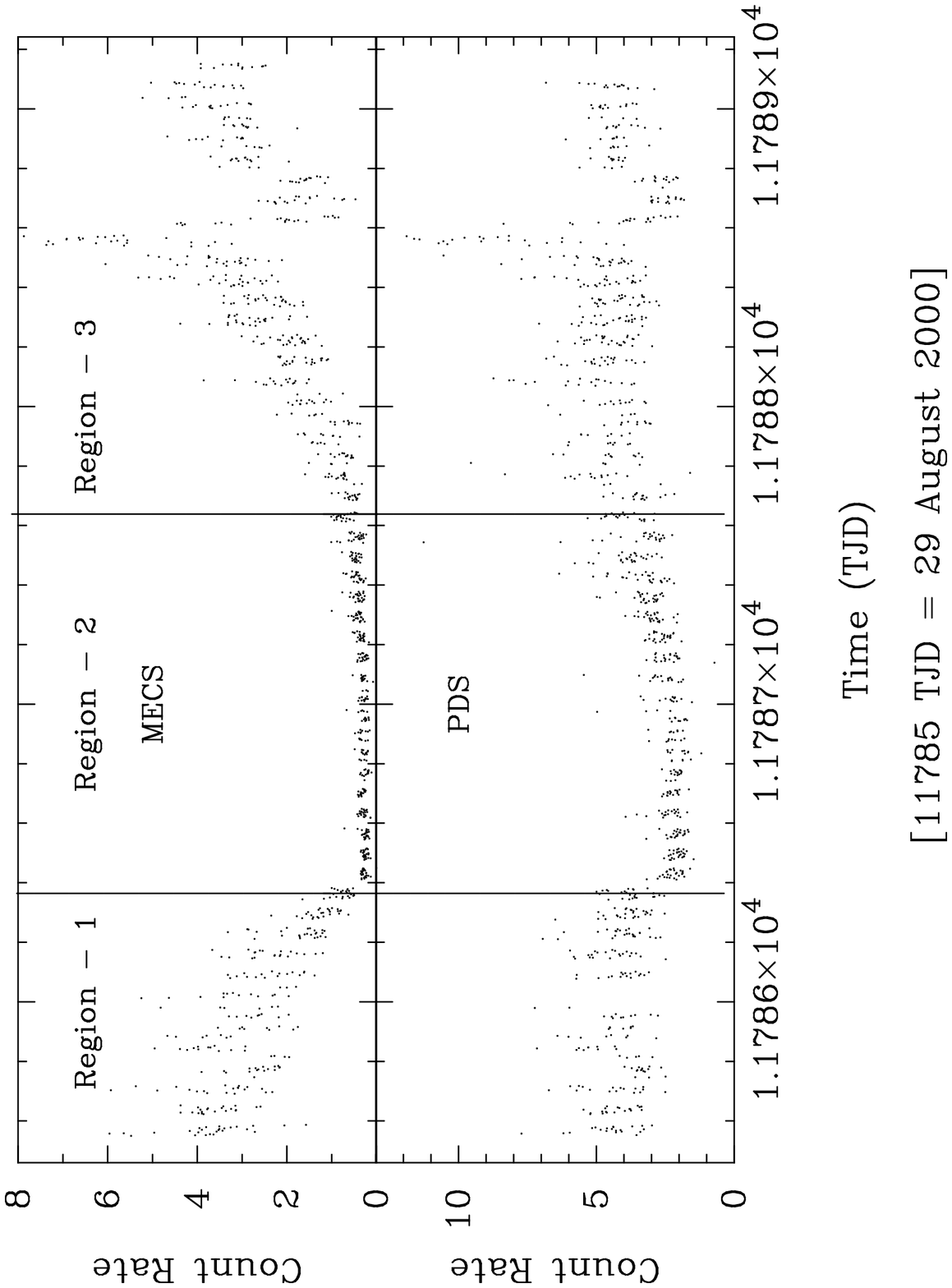}
\caption{The MECS and PDS light curves with time bin same as the spin 
period of the neutron star obtained from the 2000 \sax\ observation of
\gx.}
\label{2000lc}
\end{figure}

\gx\ suffers heavily from absorption at soft X-ray energy bands by the 
intervening cold material. We have, therefore, used data from MECS and 
PDS detectors for the timing analysis. A barycentric correction was applied 
to the arrival times of the photons. It is not possible to correct the 
arrival times for the the binary motion as the key orbital parameters 
are not known yet. A circular region of radius 4$\arcmin$ around the 
source was selected to extract light curves from MECS event data. Light 
curves in the energy bands of 1--10 keV and 15--150 keV were extracted 
from the MECS and PDS data respectively with a time resolution of 1 s.
During the 2000 observation, light curves from which are shown in 
Figure \ref{2000lc}, high count rates were detected at the beginning 
and at the end, separated by an extended low intensity state of duration 
$\sim$30 hr. Near the end of this observation, there was a flare followed 
by a shallow dip.

\begin{figure}
\centering
\vskip 8.6cm
\includegraphics{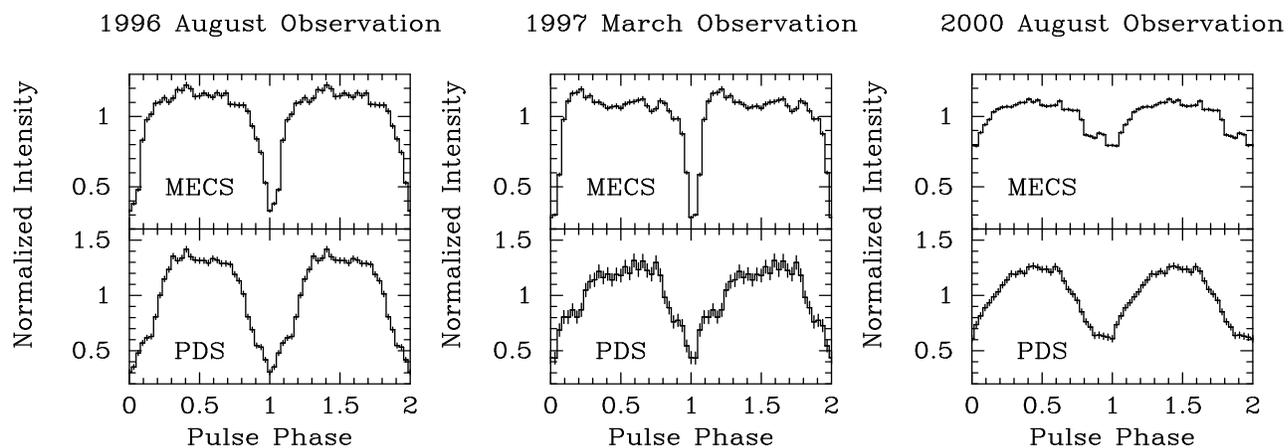}
\caption{The MECS and PDS pulse profiles of the source, of the three 
observations, are shown in top and bottom panels with 32 phase bins 
per pulse respectively. Two pulses are shown for clarity.}
\label{pp}
\end{figure}

The two minute pulses of \gx\ were visible in the raw light curves. For
accurate measurement of the pulse period, pulse folding and a $\chi^2$
maximization method was applied to all the light curves and pulse period
was determined to be 124.404(3) s, 126.018(8) s and 134.9256(10)
s during the 1996 August 18, 1997 March 25, and 2000 August 29
observations respectively. The quoted uncertainties (1$\sigma$ confidence
level) in the pulse periods represent the trial periods at which the
$\chi^2$ decreases from the peak value by one standard deviation of the
chi-sq values over a wide period range far from the peak. 
The pulse profiles, obtained from the MECS and 
PDS data of the three observations are shown in Figure~\ref{pp}. The pulse 
phases were adjusted so as to obtain the minimum in the pulse profile at 
phase zero. The characteristic narrow dips are clearly seen in the MECS 
pulse profiles of the 1996 and 1997 observations. The pulse fraction 
(defined as the ratio of the difference of maximum and minimum flux to 
the maximum flux in the pulse profile) in 1--10 keV energy band (MECS data), 
during these two observations is in the range of 70--75\%. However, the MECS 
pulse profile of the 2000 observation is different. The dip seems to be 
absent and the pulse fraction is reduced to about 30\%. Though the PDS 
pulse profiles of all three \sax\ observations look alike, the pulse 
fraction during the 2000 observation is less ($\sim$ 50\%) compared to 
the 1996 and 1997 observations (70--80\%).

\begin{figure}
\centering
\vskip 6.0cm
\includegraphics{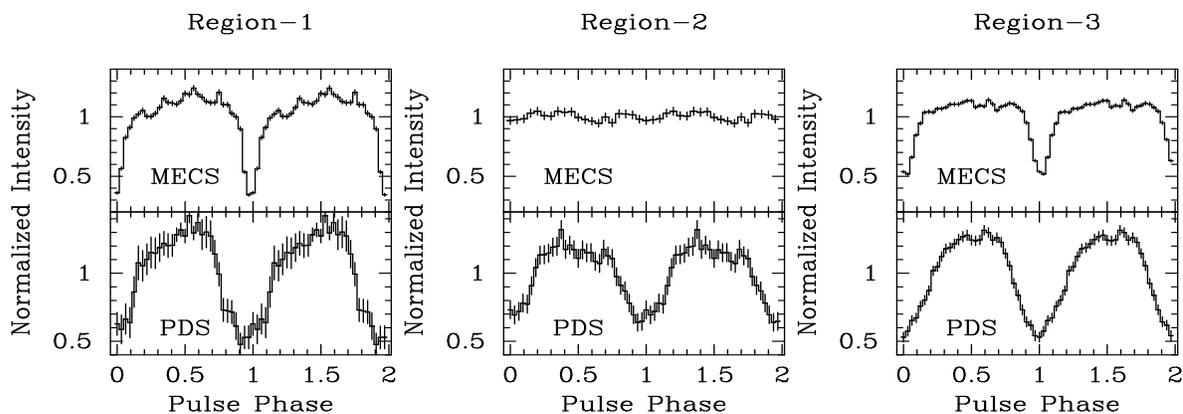}
\caption{The MECS and PDS pulse profiles of the three different regions 
of 2000 \sax\ observation (as shown in Figure~\ref{2000lc}) of \gx.}
\label{2000ef}
\end{figure}

To get a detailed picture of the pulsation properties in this extended
low state during the 2000 observation, we divided the light curve into
three different regions as shown in Figure~\ref{2000lc}. Pulsations with
profiles similar to those seen in the 1996 and 1997 observations were
detected in the MECS light curves of regions 1 and 3, whereas in region 2,
the 3$\sigma$ upper limit of soft X-ray pulse fraction is 10\%. However, 
pulsations are detected in all three regions in hard X-ray light curve of 
the PDS. The MECS and PDS pulse profiles of the three different regions are 
shown in Figure~\ref{2000ef}. 

\begin{figure}
\centering
\vskip 9.5cm
\includegraphics{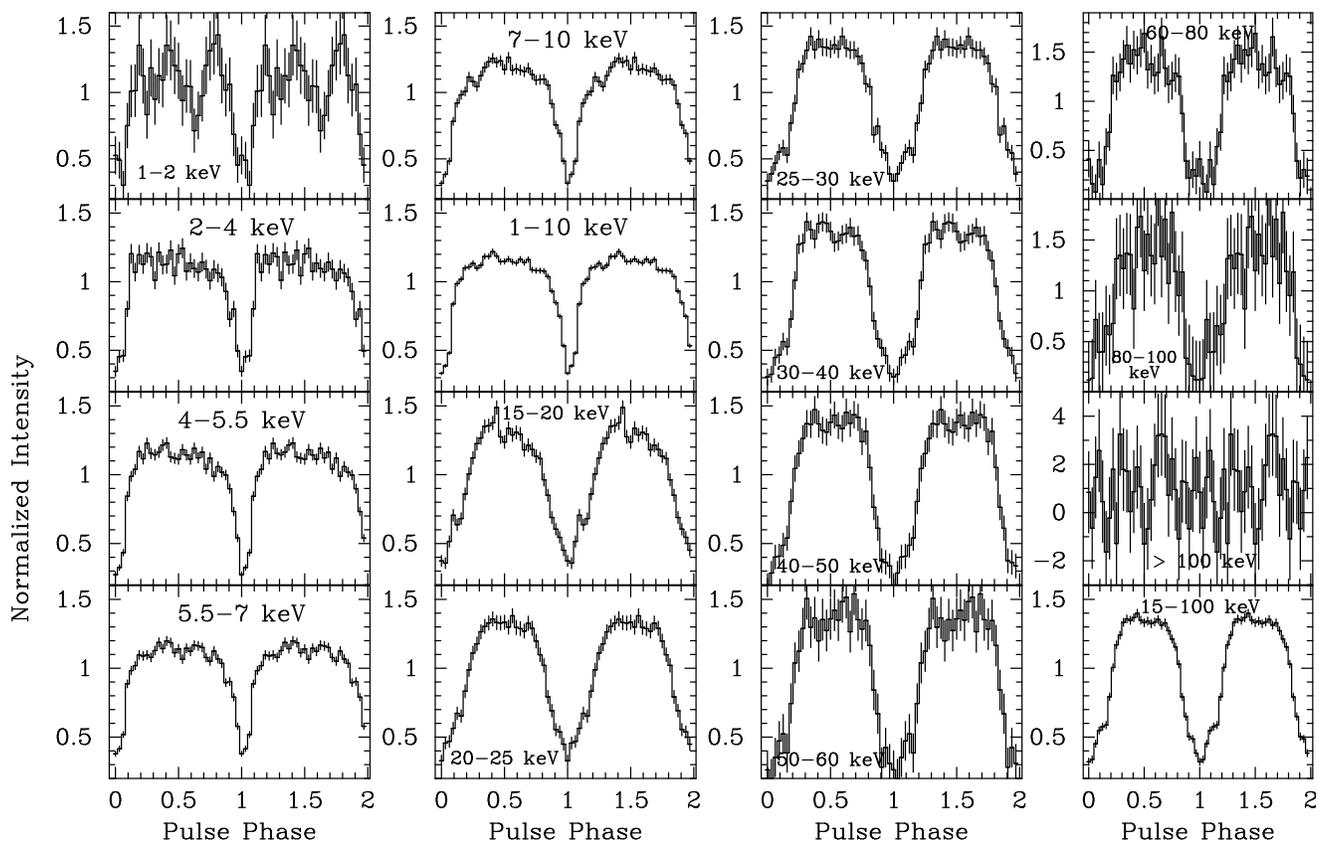}
\caption{The MECS and PDS pulse profiles of \gx\ in 16 different energy bands 
(as described in text) of 1996 August observation.}
\label{1996ef}
\end{figure}

\begin{figure}
\centering
\vskip 9.5cm
\includegraphics{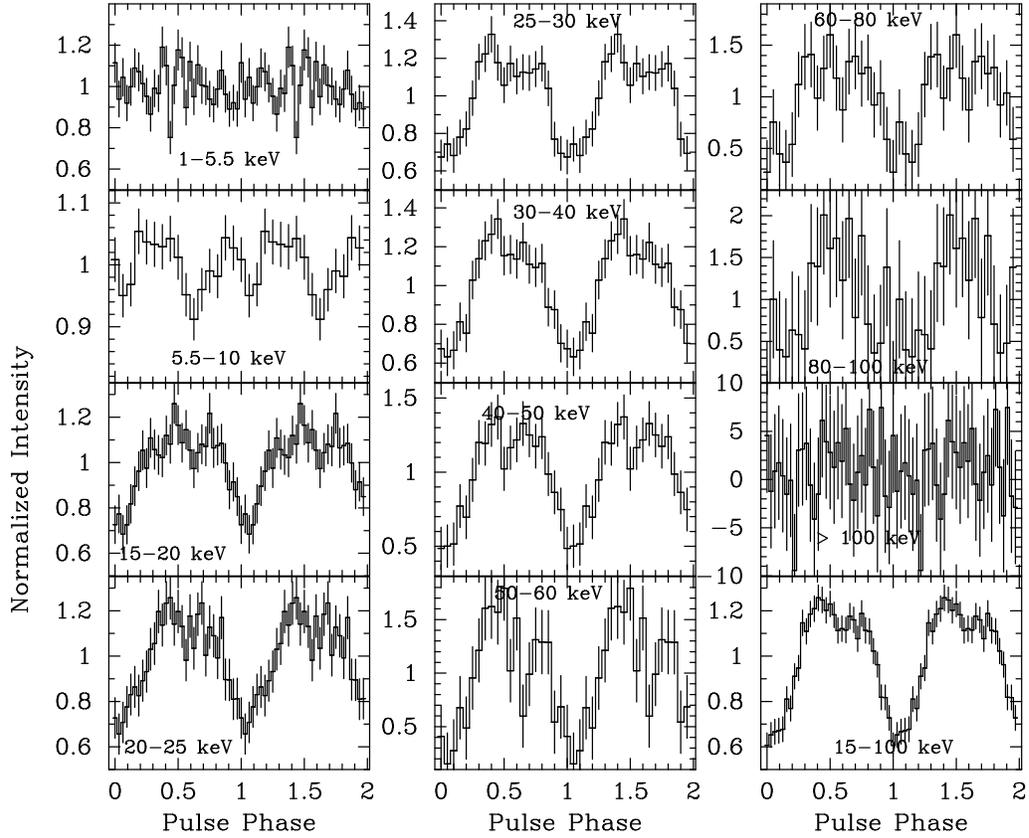}
\caption{The MECS and PDS pulse profiles of \gx\ in 12 different energy 
bands (as described in text) of region-2 of 2000 August observation.}
\label{2000r2}
\end{figure}

Figures~\ref{pp} and \ref{2000ef} show a clear difference in the shape
of pulse profiles of \gx\ in the soft and hard X-ray energy bands. The dip
is narrow in soft X-rays and gradually becomes broader with increasing energy. 
To investigate the change in the shape of pulse profiles, we generated light 
curves in 16 different energy bands from MECS and PDS event files of the 
1996 observation. The corresponding pulse profiles are shown in 
Figure~\ref{1996ef}. An increase in the width of the dips in the pulse 
profiles with energy range is apparent from the figure. It is found that 
the light curve above 100 keV is mainly background dominated and pulsations 
were not detected above 100 keV. Energy resolved pulse profiles from 
region 2 of the 2000 observation are shown in Figure~\ref{2000r2}. The 
pulse profiles in different energy bands of region 1 and 3 of 2000 August
observation are similar to those seen in 1996 and 1997 observations. But, as 
mentioned previously, pulsations are absent or much reduced in 1--5.5 keV energy 
band of region 2 of 2000 observation (Figure \ref{2000r2}). A change in pulse
shape is also noticed between the pulse profiles in 5.5--10 keV
(a double peaked profile with a pulse fraction of a mere $\sim$ 13\%)
and the same in the higher energy bands.

\section{Pulse phase averaged spectroscopy}

For spectral analysis, we have extracted LECS spectra from regions of radius
$6 \arcmin$ centered on the object for the 1997 and 2000 observations. The 
combined MECS source photons (MECS~1 was not operational in 2000)
were extracted from circular regions with a $4 \arcmin$ radius. Background
spectra for both LECS and MECS instruments were extracted from appropriate
source-free regions of the field of view by selecting annular regions around
the source. The software package named SAX Data Analysis System (SAXDAS)
was used to extract background subtracted PDS spectra from the event files
of all the three \sax\ observations. For spectral fitting, response matrices 
released by SDC in 1998 November, were used. A rebinning scheme 
(ftp://heasarc.gsfc.nasa.gov/sax/cal/responses/grouping) suggested in the 
SAX data analysis guide was used to rebin the LECS spectra to allow use of 
the $\chi^2$ statistics. Events were selected in the energy ranges 1.0--4.0 
keV for LECS, 1.65--10.0 keV for MECS and 15.0--150 keV for PDS where the 
instrument responses are well determined. Combined spectra from the LECS, 
MECS and PDS detectors, after appropriate background subtraction, were 
fitted simultaneously. All the spectral parameters, other than the relative 
normalization, were tied to be the same for all the detectors.

The spectra were first fitted to a model consisting of a power law
continuum with a Gaussian for the iron line emission and absorption by
matter along the line of sight, which yielded poor values of reduced $\chi^2$.
Following this, we fitted the spectra with a model consisting of a
Comptonized continuum along with a Gaussian function and interstellar
absorption and obtained better values for the reduced $\chi^2$. 
Addition of an absorption edge at $\sim$ 30 keV further improved 
the spectral fitting with a decrease in reduced $\chi^2$ from 1.1 to 1.0 
for region-1 and 1.50 to 1.41 for region-3, whereas the region-2 did not 
show any improvement in spectral fitting with a reduced $\chi^2$ of 2.63. 
The edge energies and the absorption depth at the threshold are found 
to be consistent (within errors) for all three \sax\ observations.
The residual around 35 keV can be fitted with either a cyclotron absorption 
feature or an absorption edge. However, as the magnetic field strength 
of the neutron star is estimated to be of the order of 10$^{13-14}$ G, the 
cyclotron absorption line is expected at higher energies ($>$ 100 keV). 
As there is no physical reason for the absorption feature at $\sim$
35 keV, the absorption edge detected here is possibly an instrumental
effect. But we note that with a similar analysis of the PDS spectrum of
Crab, we did not detect any such feature.

The equivalent width of the 6.4 keV iron K$_\alpha$ line was found to be in 
the range of 180--320 eV during the 1996, 1997 and regions 1 and 3 of the 2000 
observation. In the region 2, the equivalent width of the iron line was very high 
($\sim$ 2.9 keV) with significant line like residuals just above 7 keV. The 
region 2 spectrum, when fitted with the addition of another single Gaussian 
function gives a much smaller reduced $\chi^2$ of 1.72. We have considered 
if the second line detected most clearly in the region 2 spectrum of 2000 can be 
emission from hydrogen like iron as was found in an observation of GX~1+4 with 
the Chandra HETG (Paul et al. 2004). The 90\% confidence limit for the centre 
energy of the second line detected in the MECS spectrum of 2000 region 2 is 
7.10$\pm$0.05 keV. If we fix the center energy of the second line at 6.95 keV,
(at which the second line is detected with a 3$\sigma$ confidence level
in the Chandra spectrum) for 65 degrees of freedom the $\chi^2$ increases by 
10 compared to a line at 7.10 keV. We therefore conclude that the second emission 
line detected in the MECS spectrum is indeed K$\beta$ line of lowly ionized or 
neutral iron. Following this, we have added the second line feature in the model 
for all the spectrum. The emission line at $\sim$ 7.1 keV, with an equivalent 
width in the range of 25--80 eV is present in all \sax\ spectra of \gx\ except 
region 2 of the 2000 observation where it is very strong with an equivalent width 
of about $\sim$ 0.55 keV. 

\begin{figure}
\centering
\vskip 9.5cm
\includegraphics{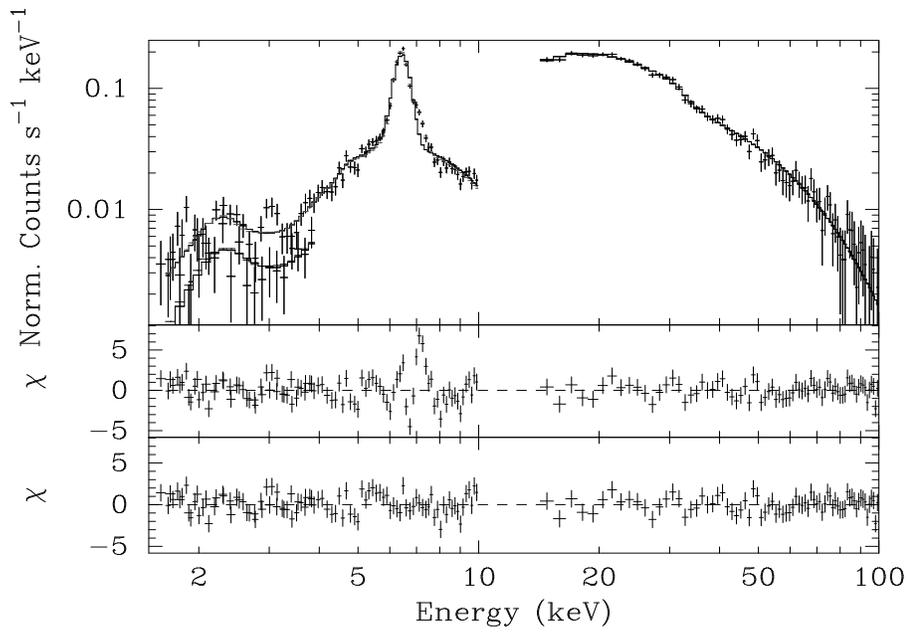}
\caption{Energy spectra of \gx\ measured with \sax\ during the region 2 of 
2000 August observation along with the fitted model (without iron $K_\beta$ 
emission line) and the residuals (middle panel). The presence of iron 
$K_\beta$ line is evident from the line like structure at $\sim$ 7.1 keV 
in the residual. Residuals obtained after including a Gaussian function at
7.1 keV in the model are shown in the bottom panel.}
\label{2000reg2}
\end{figure}

\begin{figure}
\centering
\vskip 10.3cm
\includegraphics{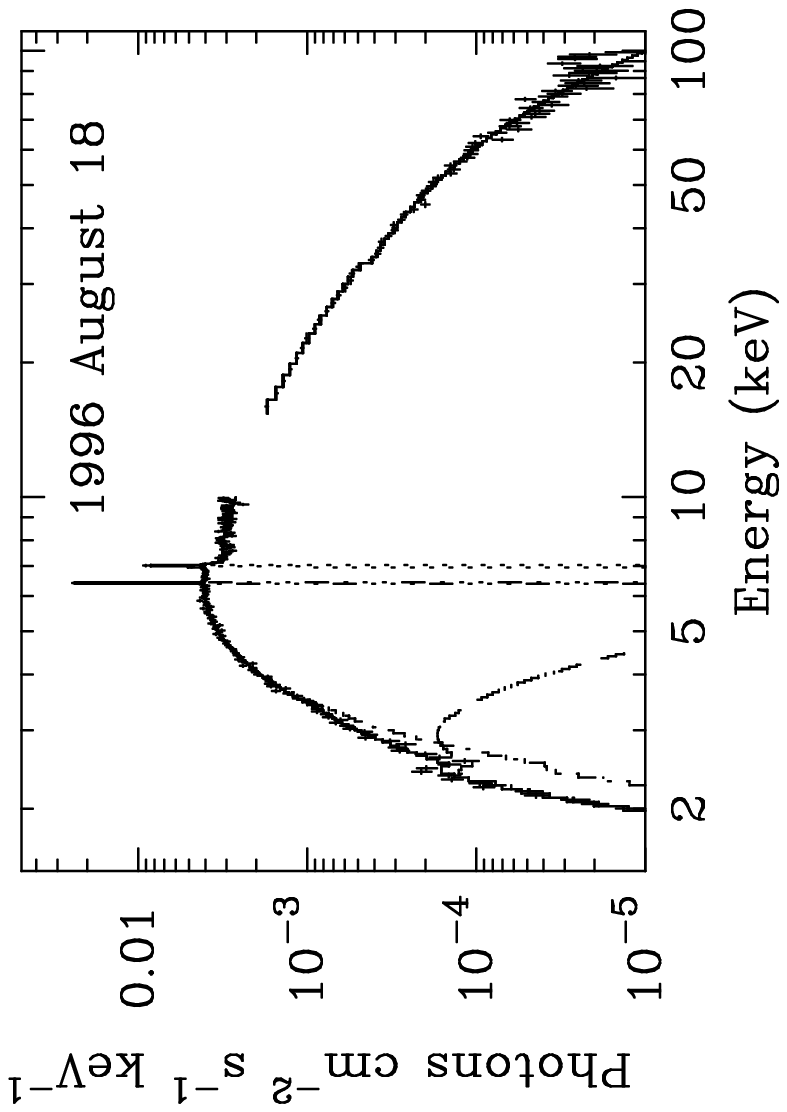}
\includegraphics{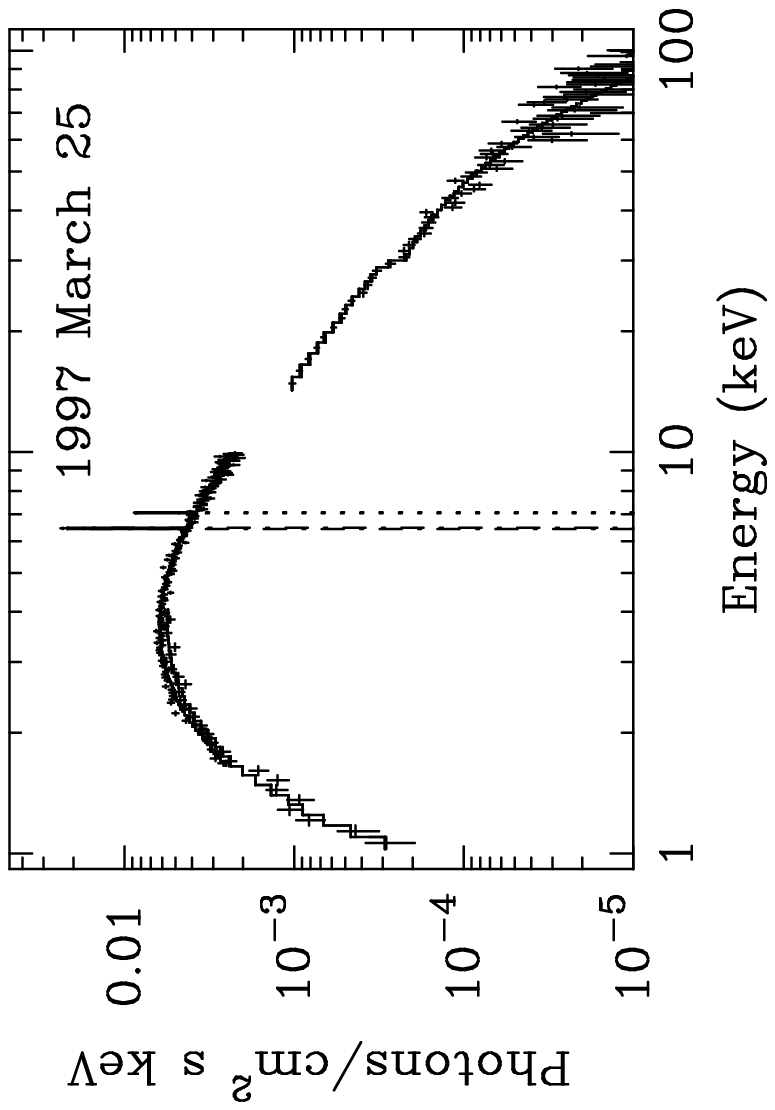}
\includegraphics{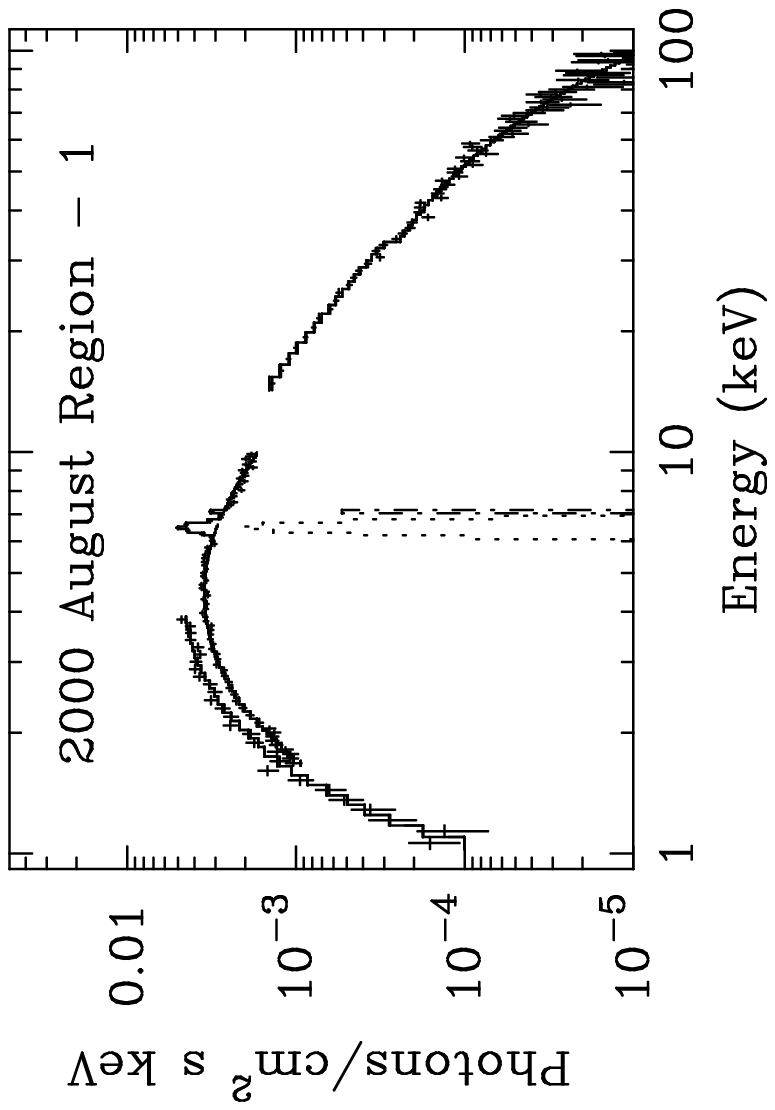}
\includegraphics{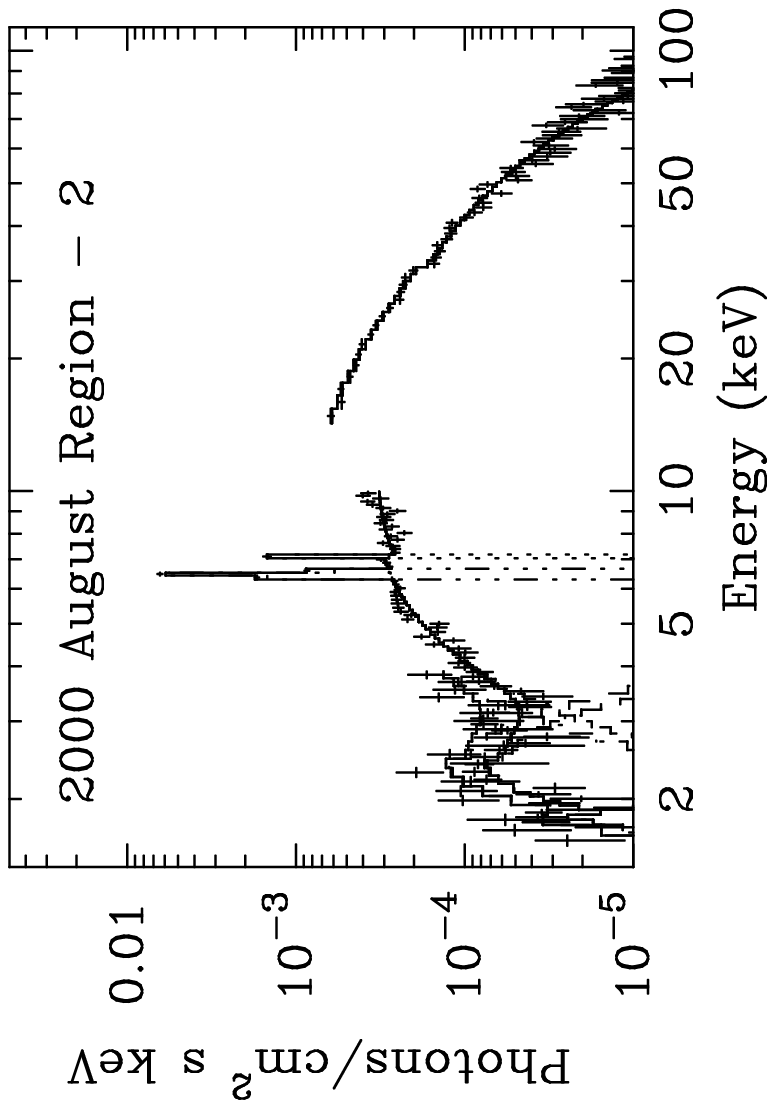}
\includegraphics{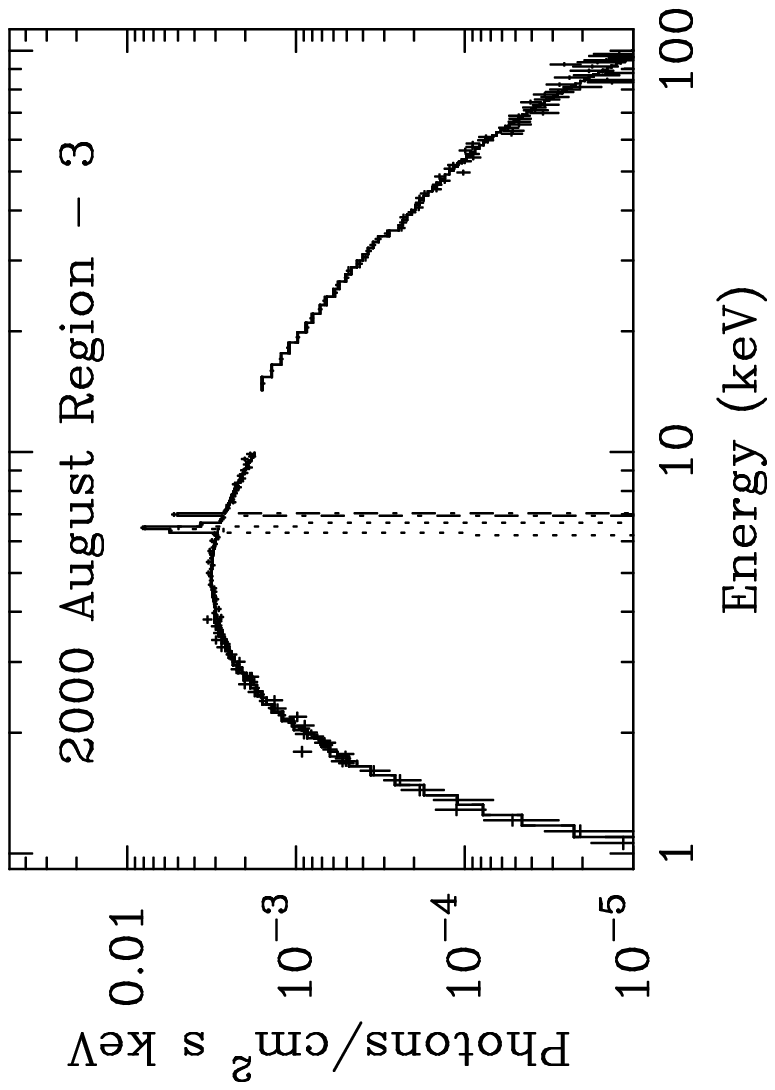}
\caption{The unfolded energy spectra of \gx\ measured with 1996 August, 1997 
March, and three regions of 2000 August \sax\ observations.}
\label{ph_av_sp}
\end{figure}

A soft X-ray excess was detected in the residuals of the spectra 
from the 1996 and region 2 of the 2000 data. We tried various model 
components such as blackbody, disk-blackbody, and bremsstrahlung to fit
this soft excess. However, addition of a bremsstrahlung component to the 
spectral model gives a better fit compared to the other two models for
the soft excess with an improved reduced $\chi^2$ of 1.11 and 1.27 for 
1996 and the region-2 of 2000 observation respectively. The integrated 
flux of the soft component in 1.0--5.0 keV energy band for 1996 and 
region-2 of 2000 observations is estimated to be 
%1.12${^{+1.3}_{-0.9}}$ $\times$ 10$^{-12}$ ergs cm$^{-2}$ s$^{-1}$ and 
%4.70${^{+28.7}_{-0.5}}$ $\times$ 10$^{-13}$ ergs cm$^{-2}$ s$^{-1}$ which 
1.12 $\times$ 10$^{-12}$ ergs cm$^{-2}$ s$^{-1}$ and 4.7 $\times$ 10$^{-13}$ 
ergs cm$^{-2}$ s$^{-1}$ which is $\sim$ 4\% and 18\% of the total model flux
in above energy band respectively. The 3$\sigma$ lower limit of above two
observations are 2.2 $\times$ 10$^{-13}$ ergs cm$^{-2}$ s$^{-1}$ and 4.2 
$\times$ 10$^{-13}$ ergs cm$^{-2}$ s$^{-1}$ respectively whereas the 3$\sigma$ 
upper limit of the soft X-ray flux during 1997, region-1 and region-3 of 2000 
observations are 6.3 $\times$ 10$^{-13}$, 4.1 $\times$ 10$^{-13}$ and 1.6 $\times$ 
10$^{-13}$ ergs cm$^{-2}$ s$^{-1}$ respectively. We did not try the non-solar 
abundance for the soft component as it is seen only in 1996 and region-2 of 2000 
observations and not in the others. Therefore, the detected soft excess in 
\gx\ is unlikely to be due to abundance anomaly. The spectral parameters 
with 3$\sigma$ error estimates and the reduced $\chi^2$ obtained are given 
in Table~\ref{par}. The region-2 spectrum along with the best fit model 
components and residuals without and with the iron K$_\beta$ line are shown 
in Figure~\ref{2000reg2}. The unfolded spectra measured during 1996, 1997 
and 2000 \sax\ observations are shown in Figure~\ref{ph_av_sp}.

\section{Pulse phase resolved spectroscopy}

Since the dip in the pulse profile of GX~1+4 shows significant energy
dependence, pulse phase resolved spectroscopy may hold a clue to the
origin of the dip. To estimate the change in spectral parameters during the
dips, we have carried out pulse phase resolved spectroscopy during all
the observations. The LECS and MECS spectra were accumulated into 10 
pulse phase bins by applying phase filtering in the FTOOLS task XSELECT and 
SAXDAS software package was used to extract PDS spectra at the same pulse 
phases. As in the case of phase-averaged spectroscopy, the background 
spectra were extracted from source free regions in the event files and 
appropriate response files were used for the spectral fitting. The LECS, 
MECS and PDS spectra of region 1 of 2000 observation were added with the 
corresponding spectra of region 3 to improve the statistics. 

For the phase resolved spectral analysis, initially all the spectral
parameters were allowed to vary. However, we did not find any systematic
variation in the value of $N_H$ and iron emission line parameters with
the pulse phase. We tried to fit the phase resolved spectra of all three
\sax\ observations by freezing $N_H$, the absorption edge and iron line 
parameters, which did not show any variation over pulse phase, to the phase 
averaged values. We included the Bremsstrahlung component to the spectral 
fitting for the 1996 and 2000 region-2 phase resolved spectra as it was 
present in the phase averaged spectra. It was found that the change in 
continuum flux with pulse phase in 2--10 keV and 10--100 keV energy ranges 
is consistent with that of the MECS and PDS pulse profiles respectively. 
We found a marginal change in the values of the input spectrum temperature 
$kT_0$, plasma temperature $kT$ and the scattering optical depth $\tau$ over
pulse phase in three \sax\ observations. The results of 1996, 1997 and 
combined regions of 1 \& 3 of 2000 observations are found to be similar. It 
is found that the results obtained from the phase resolved spectroscopy of 
region-2 of 2000 observation are similar to that obtained by Galloway et al. 
(2001) e.g. an increase in the optical depth at the pulse minimum.

\section{Discussion}

\subsection{Pulsations during the extended low state}

On several occasions, the accretion powered X-ray pulsar \gx\ was found
to show a decrease in X-ray luminosity by an order of magnitude compared to
the flux before and after such a state. The low state lasts for at least
a few hours and can have rapid ingress to and egress from the low state within
only about 4-6 hours (Giles et al. 2000; present work). Many short exposures 
with RXTE-PCA also found the source at a very low flux level, about an order 
of magnitude lower than usual, (Cui 1997, Cui \& Smith 2003). During these low 
flux episodes, the pulsations were either absent, or the pulse fraction was 
reduced significantly from the usual 50-80\%, often below the detection limit 
of RXTE-PCA. This has been interpreted as due to onset of centrifugal barrier 
or the propeller effect. With decrease in the mass accretion rate, when the 
inner radius of the accretion disk recedes beyond the co-rotation radius of the 
neutron star, this effect may set in. To compare the 2--10 keV flux during the 
low-state reported here with the same during earlier reported low intensity
episodes with the RXTE-PCA, when the pulsations became undetectable,
we have reanalyzed the RXTE-PCA data of those observations. The 2--10 keV
flux measured from these observations are shown in Table 2.

During the low state reported here, the absorbed X-ray flux in 
2--10 keV band was 2.9$\times$ 10$^{-11}$ ergs cm$^{-2}$ s$^{-1}$ and 
is comparable to the earlier episodes when pulsations were below the detection 
level of the RXTE. However, we found that as the X-ray spectrum of GX~1+4 is
very hard, the total X-ray luminosity in 10--100 keV energy band during this 
state is more than an order of magnitude higher than that in the 2-10 keV band 
and pulsations are clearly detected in the hard X-rays. The spectral analysis 
of the \sax\ data showed that in the low state, the medium energy (2--10 keV)
X-ray flux is reduced by an order of magnitude, mostly due to an increase
in the absorption. In comparison, the hard X-ray flux is reduced by a
factor of two only. It was also pointed out by Cui (1997) from RXTE-PCA
data that the X-ray spectrum is harder in the low state. Therefore,
there can be significant flux in the hard X-ray band, where the effective
area of the RXTE-PCA detectors fall rapidly. It is possible
that the low states reported by Cui (1997) and Cui and Smith (2004) are
similar to the one presented here. Only short exposures and decreasing
hard X-ray effective area of the RXTE-PCA did not allow detection of the
hard X-ray pulses in RXTE-PCA observations during the low states. In such 
a scenario, the propeller regime for GX~1+4 may occur at a still lower
mass accretion rate than that reported earlier.

\subsection{Broad band X-ray spectrum and its changes during the
extended low-state}

From the three \sax\ observations reported here, the pulse-phase-averaged broad
band X-ray spectrum in both high and low states of GX~1+4 is found to agree
well with a Comptonization model with additional components for absorption,
line emissions, and in some cases, a soft excess. During the extended low 
state in 2000, the continuum flux in the 2--10 keV decreased by a factor  
of $\sim$10 whereas the 10--100 keV flux decreased only by about 50\%.
An intense iron emission line at 6.4 keV is clearly seen in the phase
averaged spectra in both the high and low intensity states. In addition,
a strong emission line at 7.1 keV was also detected in the low state spectrum,
which was very weak during the high intensity states. The $K_\alpha$ line
flux was about 3--5 times larger than the $K_\beta$ line flux in all the
observations and is consistent with the ratio expected for fluorescence
emission from neutral iron. With a moderate energy resolution of MECS it
is not possible to determine the exact ionization state of the reprocessing
material.

We have also examined changes of the two hardness ratios
(1.0--4.0/7.0--10.0 keV and 4.0-7.0/7.0--10.0) in the MECS energy bands
during the \sax\ observation in 2000. The hardness ratios clearly
show increased absorption during the low state. There was no significant
spectral change associated with the intensity variations near the
end of this observation.

Although a similar low state of \gx\ was observed earlier, the event
lasted for only about 6 hours and detailed spectroscopy was not feasible
due to poorer photon statistics and lower energy resolution of the RXTE-PCA
(Galloway et al. 2000). In this paper, we present the first detailed
spectroscopy of \gx\ in this state. Though both the soft and hard X-ray
flux decreased during the low state, the optical depth and temperature of
the Comptonizing plasma are comparable (within errors) with those during
the high intensity states just before and after the low intensity episode.
The absorption column density is an order of magnitude higher during the
low state. An increase in absorption may not be sole reason for the low
state as the hard X-ray flux also decreased by about 50\%, and the
temperature of the seed photons increased during the low state by
a factor of about two. However, the decrease in the hard X-ray flux
may arise due to Thompson scattering from the line of sight if the absorbing 
material is partially ionized and the actual material in the line of sight is
more than that measured with the simple absorber model. In the 
Comptonization model for X-ray pulsar spectra, the seed photons
are likely to be produced from a heated surface of the neutron star. If
there is a spherical component of accretion, it may cause heating of the
surface and these seed photons undergo Compton upscattering in the accretion
column. An increase in the temperature of the seed photons during the low
state also indicates that the low state is not only due to an increased
absorption along the line of sight. There must have been some additional
change in the accretion process, for example, and enhancement in the
spherical component of mass accretion.

Fluorescence emission lines with very high equivalent width similar to those
found in \gx\ in the low state are seen in some persistent X-ray binary pulsars 
like Her~X-1, LMC~X-4 and SMC X-1 during their low-intensity phases of the
super orbital period (Naik \& Paul 2003; Naik \& Paul 2004, Vrtilek et al.
2001). In these systems, the low-states arise due to increased absorption 
of the X-ray continuum by precessing warped inner accretion disk, resulting 
in a very large equivalent width. However, unlike these systems, in \gx\ the 
extended low state is observed intermittently, without any hint of
a periodicity. The RXTE/ASM light curve of \gx\ does not show any 
periodic or quasi-periodic long term intensity variation as seen in Her~X-1, 
LMC~X-4 and SMC~X-1. The difference in the duration and absence of 
periodicity in the occurrence of the low states of \gx\ rules out a similar 
mechanism for this source. 

\subsection{Spectral features of the absorption dip in pulse profiles of \gx}

Energy resolved pulse profiles of \gx\ obtained from \sax\ observations
provide useful information towards understanding the accretion geometry of the
binary system. The characteristic absorption dips are detected in wide energy 
range (1--100 keV). The pulse profiles in narrow energy bands as presented 
here in this paper, show a gradual change with energy. An increase in width 
of the absorption dip with energy is clearly seen in all three \sax\ 
observations. These narrow and sharp dips have been attributed to the 
interaction between the accretion column and the emission region of the 
neutron star (Galloway et al. 2001). Partial eclipses of the emission region 
by the accretion column occurring during the spinning of the neutron star are 
reflected as dips in the pulse profile. The closest approach of the accretion
column to the line of sight corresponds to the minimum intensity in the pulse
profile. Similar sharp and narrow dips are observed in the pulse profiles of
two other X-ray pulsars RX~J0812.4--3114 (Reig \& Roche 1999) and A~0535+262
(Cemeljic \& Bulik 1998). However, a gradual increase in the width of the
absorption dip with energy is seen only in \gx. A widening of the dip at 
higher energies may result from changes in the beam pattern with energy. 
Outside the dip, the high energy photons are typically those which have 
undergone many scatterings. Except along the accretion column, where both 
the low and high energy photons are strongly suppressed, the high energy 
photons escape from the column preferentially at large angles, whereas 
low-energy photons are more isotropic. The emission geometry may change 
from a pencil beam at low energy to a fan beam at high energy.

We have detected marginal evidence of spectral changes during the pulse phase
in the form of a varying optical depth, plasma temperature and temperature
of the soft seed photons. The pulse phase dependence of these parameters is 
not identical during different observations and the implication of the 
pattern of these variations is not clear.

\begin{deluxetable}{llllll}
\footnotesize
\tablecaption{\sax\ observations of \gx}
\tablewidth{0pt}
\tablehead{
\colhead{Year of } &\colhead{Start Time} &\colhead{End Time} &\colhead{LECS Exp.} &\colhead{MECS Exp.} &\colhead{PDS Exp.}\\
\colhead{Observation}  &\colhead{(Date, UT)}   &\colhead{(Date, UT)}  &\colhead{(ks)}  &\colhead{(ks)} &\colhead{(ks)}}
\startdata
1996  &August 18, 06:11  & August 19, 03:38  &-----  &38.6   &17.6\\
1997  &March 25, 22:43   & March 26, 16:08   &13     &31.5   &13.5\\
2000  &August 29, 12:36  & September 02, 03:38  &56.5  &132  &58.5\\
\enddata
\label{obs}
\end{deluxetable}

\begin{deluxetable}{llllll}
\footnotesize
\tablecaption{Spectral parameters for \gx\ during different intensity states}
\tablewidth{0pt}
\tablehead{
\colhead{Parameter} &\colhead{1996 August} &\colhead{1997 March} &\colhead{2000 August} &\colhead{2000 August} &\colhead{2000 August}\\
\colhead{}  &\colhead{}   &\colhead{}  &\colhead{Region--1}  &\colhead{Region--2} &\colhead{Region--3}}
\startdata
N$_H$$^1$        &21$\pm$2  &1.1$\pm$0.1 &1.37$^{+0.05}_{-0.15}$   &14.8$^{+1.0}_{-3.4}$  &2.4$\pm$0.1\\
$kT_{Br}$ (keV)  &0.36$^{+0.12}_{-0.11}$   &------   &------  &0.26$^{+0.06}_{-0.04}$  &------\\
$T_0$ (keV)	 &1.48$^{+0.07}_{-0.06}$   &1.28$^{+0.04}_{-0.05}$  &1.62$^{+0.04}_{-0.02}$ &3.5$^{+0.6}_{-0.4}$  &1.73$\pm$0.03\\
$kT_{Co}$ (keV)  &13$^{+0.4}_{-0.5}$  &12.7$^{+1.2}_{-1.1}$ &15.2$^{+0.5}_{-1.6}$  &12.6$^{+1.9}_{-0.9}$  &14.3$\pm$0.6\\
$\tau$           &3.04$^{+0.19}_{-0.14}$   &3$^{+0.4}_{-0.3}$  &2.38$^{+0.38}_{-0.09}$ &3.1$^{+0.3}_{-0.5}$ &2.55$\pm$0.13\\
\\
Iron K$_\alpha$ emission line \\
\\
Fe$_E$ (keV)$^2$ &6.42$\pm$0.02    &6.45$^{+0.03}_{-0.02}$    &6.49$\pm$0.03 &6.47$\pm$0.01  &6.45$\pm$0.02\\
Line width (keV) &0.1$\pm$0.02     &0.05$^{*}$    &0.12$\pm$0.06  &0.05$^{*}$ &0.07$\pm$0.05\\
Eqw. width (keV) &0.22$\pm$0.01 &0.19$\pm$0.02   &0.24$\pm$0.03  &3.0$^{+0.6}_{-0.3}$  &0.33$\pm$0.03\\
Line Flux$^3$    &1.31$^{+0.07}_{-0.10}$   &0.85$\pm$0.08  &0.77$\pm$0.09 &1.08$^{+0.21}_{-0.12}$  &1.04$\pm$0.09\\
\\
Iron K$_\beta$ emission line \\
\\
Fe$_E$ (keV)$^2$ &6.99$^{+0.30}_{-0.05}$    &7.06$^{+0.10}_{-0.07}$    &7.1$^\dag$ &7.10$^{+0.02}_{-0.05}$ &7.01$^{+0.08}_{-0.09}$\\
Eqw. width (keV) &0.06$\pm$0.01  &0.06$\pm$0.02  &0.02$\pm$0.02 &0.55$^{+0.14}_{-0.09}$ &0.08$^{+0.04}_{-0.02}$\\
Line Flux$^3$    &0.38$\pm$0.08  &0.25$^{+0.10}_{-0.08}$ &0.05$^{+0.10}_{-0.01}$  &0.23$^{+0.06}_{-0.04}$  &0.25$^{+0.12}_{-0.07}$\\
\\
\\
E$_{edge}$ (keV) &33$^{+1}_{-2}$  &30$\pm$3 &33$^{+2}_{-3}$ &32$^{+13}_{-8}$ &35$\pm$2\\
Absorption depth  &0.13$^{+0.03}_{-0.05}$   &0.28$^{+0.12}_{-0.11}$   &0.19$^{+0.11}_{-0.07}$   &0.11$^{+0.11}_{-0.06}$  &0.15$\pm$0.06\\
Reduced $\chi^2$  &1.18 (239 dof)  &1.08 (271 dof)  &1.0 (168 dof) &1.27 (158 dof) &1.23 (168 dof)\\
\\
Model Flux$^4$\\
\\
2--10 keV   &2.4$^{+0.3}_{-0.2}$  &3.2$^{+0.5}_{-0.3}$   &2.07$^{+0.34}_{-0.08}$  &0.29$^{+0.08}_{-0.05}$  &2.01$^{+0.13}_{-0.15}$\\
10--100 keV &14.9$^{+1.1}_{-0.9}$ &7.3$^{+0.8}_{-1.1}$   &10.1$^{+1.6}_{-0.3}$  &5.1$^{+1.1}_{-0.9}$ &11.2$^{+0.7}_{-0.8}$\\
%Unabsorbed soft excess flux  &14.23$^{+2.66}_{-13.05}$  &----- &----- &7.18$^{+19.70}_{-5.44}$  &-----\\
\enddata
\tablenotetext{}{The errors given here are for 90\% confidence limit.  $^1$ : 10$^{22}$ atoms cm$^{-2}$, $^2$ : Iron emission line energy, $^3$ : 10$^{-11}$ ergs 
cm$^{-2}$ s$^{-1}$, $^4$ : 10$^{-10}$ ergs cm$^{-2}$ s$^{-1}$}
\tablenotetext{}{Fitted Model = Wabs (Br + CompTT + Ga1 + Ga2) Edge. 
Wabs = Photoelectric absorption parameterized as equivalent hydrogen column 
density N$_H$, Br = thermal-bremsstrahlung-type component with plasma 
temperature $kT_{Br}$, CompTT = thermal Comptonized component, and Ga1 
\& Ga2 = Gaussian function for iron K$_\alpha$ and K$_\beta$ emission lines. 
$^{*}$ : Upper limit to the line width, $^\dag$ : Parameter kept fixed.}
\tablenotetext{}{\\ 2--10 keV flux during the low state observed in GX~1+4
with RXTE on various occasions are \\
\\1996 July 20 RXTE Observation (Giles et al. 2000)  = 0.35 $\times$ 10$^{-10}$ ergs cm$^{-2}$ s$^{-1}$ \\
1996 September 25 RXTE observation (Cui 1997) = 0.40 $\times$ 10$^{-10}$ ergs cm$^{-2}$ s$^{-1}$ \\
2002 July 2 RXTE observation (Cui \& Smith 2004) = 0.23 $\times$ 10$^{-10}$ ergs cm$^{-2}$ s$^{-1}$ }
\label{par}
\end{deluxetable}

\section*{Acknowledgments}
We thank an anonymous referee for suggestions which
helped us to improved the content of this paper.
The Beppo-SAX satellite is a joint Italian and Dutch program.
We thank the staff members of Beppo-SAX Science Data Center and
RXTE/ASM group for making the data public. SN is supported by
IRCSET through EMBARK postdoctoral fellowship.

{}

%\begin{figure}
%\centering
%\vskip 11.4cm
%\special{psfile=f14.eps vscale=75 hscale=80 hoffset=-145 voffset=445 angle=-90}
%\special{psfile=f15.eps vscale=75 hscale=80 hoffset=95 voffset=445 angle = -90}
%\caption{The change in the spectral parameters of \gx\ obtained from the
%broad-band pulse-phase-resolved spectroscopy of the LECS, MECS, and PDS 
%spectra during the region 2 (left panel) and combined regions 1 and 3 
%(right panel) of 2000 August observation.}
%\label{ph_rs2}
%\end{figure}


\begin{thebibliography}{}
\bibitem[]{} Boella, G., Butler, R. C., Perola, G. C., et al. 1997, A\&AS, 122, 299
\bibitem[]{} Cemeljic, M., \& Bulik, T. 1998, Acta Astron., 48, 65
\bibitem[]{} Chakrabarty, D., Bildsten, L., Finger, M. H., et al. 1997, ApJ, 481, L101
\bibitem[]{} Chakrabarty, D. \& Roche P. 1997, ApJ, 489, 254
\bibitem[]{} Cui, W. 1997, ApJ, 482, L163
\bibitem[]{} Cui, W. \& Smith, B. 2004, ApJ, 602, 320
\bibitem[]{} Dotani, T, Kii, T., Nagase, F., et al. 1989, PASJ, 41, 427
\bibitem[]{} Frontera, F., Costa, E., Dal Fiume, D., et al. 1997, A\&AS, 122, 357
\bibitem[]{} Galloway, D. K. 2000, ApJ, 543, L137
\bibitem[]{} Galloway, D. K., Giles, A. B., Greenhill, J. G., \& Storey, M. C. 2000, MNRAS, 311, 755
\bibitem[]{} Galloway, D. K., Giles, A. B., Wu, K., \& Greenhill, J. G. 2001, MNRAS, 325, 419
\bibitem[]{} Ghosh, P., \& Lamb, F. K. 1979, ApJ, 234, 296
\bibitem[]{} Giles, A. B., Galloway, D. K., Greenhill, J. G., et al. 2000, ApJ, 529, 447
\bibitem[]{} Greenhill, J. G., Sharma, D. P., Dieters, S. W. B. et al. 1993, MNRAS, 260, 21 
\bibitem[]{} Makishima, K., Ohashi, T., Sakao, T., et al. 1988, Nature, 33, 746
\bibitem[]{} Naik, S., \& Paul, B. 2003, A\&A, 401, 265
\bibitem[]{} Naik, S., \& Paul, B. 2004, ApJ, 600, 351
\bibitem[]{} Paul, B., Rao, A. R., \& Singh, K. P. 1997a, A\&A, 320, L9
\bibitem[]{} Paul, B., Agrawal, P. C., Rao, A. R., \& Manchanda, R. K. 1997b, A\&A, 319, 507
\bibitem[]{} Paul, B., Dotani, T., Nagase, F., Mukherjee, U., \& Naik, S. 2004, ApJ (submitted)
\bibitem[]{} Pereira, M. G., Braga, J., \& Jablonski, F. 1999, ApJ, 526, L105
\bibitem[]{} Rao, A. R., Paul, B., Chitnis, V. R., Agrawal, P. C., \& 
Manchanda, R. K. 1994, A\&A, 289, L43
\bibitem[]{} Reig, P., \& Roche, P. 1999, MNRAS, 306, 95
\bibitem[]{} Vrtilek, S. D., Raymond, J. C., Boroson, B., Kallman, T.,
    Quaintrell, H., \& McCray, R.  2001, \apj, 563, L139
\end{thebibliography}
\end{document}